\documentstyle[12pt]{article}
\def\co{{\cal O}}
\def\bkhat{\widehat{B}_K}
\def\NDR{{\rm NDR}}
\def\CONT{{\em cont}}
\def\LATT{{\em lat}}
\def\frac#1#2{{\textstyle{#1\over#2}}}
\def\Tr{\mathop{\rm Tr}\nolimits}
\def\bar#1{\overline{#1}}
\def\chibar{\bar\chi}

\def\gam#1{\overline{(\gamma_{#1}\otimes I)} }
\def\iden{\overline{(I\otimes I) } }
\def\sfno#1#2{\overline{(\gamma_{#1}\otimes\xi_{#2})}}

\begin{document}
\input{psfig}
%----------------------------------------------------------------------%
\begin{titlepage}
 \null
 \begin{center}
  \makebox[\textwidth][r]{UW/PT-93-25}		% for uw
 \vskip1.in
  {\Large $B_K$ USING STAGGERED FERMIONS: \\[.5em] AN UPDATE}
  \par
 \vskip 2.0em 
 {
  \begin{tabular}[t]{c}
\large    Stephen R. Sharpe \footnotemark\\[1.em]
	\em Physics Department, FM-15 \\
	\em University of Washington \\
	\em Seattle, WA 98195
  \end{tabular}}
 \par \vskip 3.0em
	 {\large\bf Abstract}
  \end{center}
\quotation
Improved results for $B_K$ are discussed.
Scaling corrections are argued to be of $O(a^2)$, leading to
a reduction in the systematic error.
For a kaon composed of degenerate quarks, the quenched result is
$\bkhat= 0.825 \pm 0.027 \pm 0.023$.
\endquotation
\vspace{1in}
\begin{center}
   {\em Poster presented at {\em LATTICE 93}, \\
	International Symposium on Lattice Field Theory, \\
	Dallas, Texas, U.S.A. 12-17 October, 1993. \\
	To appear in the proceedings.}
\end{center}
\footnotetext{email: sharpe@galileo.phys.washington.edu}
\vfill
\mbox{November 1993}
\end{titlepage}
%----------------------------------------------------------------------%
\section{INTRODUCTION}

I discuss here the improvements made by the ``staggered''
group (Gupta, Kilcup, Patel and myself) since LATTICE 91
in the calculation of
\begin{equation}
\label{bkeqn}
B_K(a) = {\langle{\overline{K}|\bar s \gamma_\mu(1\!+\!\gamma_5) d\
                         \bar s \gamma_\mu(1\!+\!\gamma_5) d|K}\rangle
	  \over {8\over3}
          \langle{\overline{K}|\bar s \gamma_\mu\gamma_5 d|0}\rangle
	  \langle{     0|\bar s\gamma_\mu\gamma_5 d|K}\rangle } \ .
\end{equation}
In 1991 I quoted results using lattices at $\beta=6$, $6.2$ and $6.4$
\cite{sharpelat91}.
The major uncertainties were:
\begin{itemize}
\item
%\noindent $\bullet$
Linear vs. quadratic extrapolation in the lattice spacing $a$.
These two choices led to $\bkhat=B_K \alpha_s^{-2/9}=0.66(6)$ and $0.78(3)$
respectively, with the errors being statistical.

\item
%\noindent $\bullet$
Perturbative corrections were not included, because
of the uncertainty in the choice of $g$.

\item
%\noindent $\bullet$
Contamination from excited states.

\item
%\noindent $\bullet$
Quenching, although there was preliminary evidence
that this was not the dominant source of error \cite{kilcuplat90,fukulat91}.

\item
%\noindent $\bullet$
Extrapolation from degenerate quarks 
($m_1=m_2\approx m_s/2$) to the physical kaon 
($m_1=m_s$, $m_2\approx 0$). From the quenched data, we estimated
that this extrapolation increased $B_K$ by 3\%,
an estimate included in our results for $\bkhat$.
The error in this estimate was, however, large.
\end{itemize}

A number of these uncertainties have now been substantially reduced,
both due to our work and that of the Kyoto-Tsukuba group.

\section{SCALING VIOLATIONS ARE $O(a^2)$}

According to the standard lore, the staggered fermion action is
good to $O(a^2)$ (up to logarithms, which will always be kept implicit).
I sketched the perturbative argument for this in Ref. \cite{sharpelat91},
and presented some supporting numerical evidence from the spectrum.
By contrast, for matrix elements of external operators
the corrections are expected to be of $O(a)$.
For example, the operators appearing in both numerator
and denominator of $B_K$ (Eq. \ref{bkeqn})
have $O(a)$ terms in their tree-level perturbative matrix elements \cite{PS}.
These $O(a)$ terms turn out to have the wrong flavor 
to contribute to the matrix element in $B_K$, 
but in 1991 it seemed possible that terms of $O(g^{2n}a)$, from n-loop
diagrams, might contribute.

In the following I sketch an argument that they do not \cite{sharpeprep}:
the scaling corrections to the matrix elements in $B_K$ are of $O(a^2)$
to all orders in perturbation theory.
The argument is an application of Symanzik's perturbative improvement 
program \cite{symanzik},
%This program has only been fully justified for scalar field theories,
%but there is no reason to expect it to break down for gauge theories or
%fermions. It has been used extensively for Wilson fermions.

It is useful to begin by demonstrating that the staggered action is
already ``improved'', i.e. has no corrections of $O(a)$.
In the notation of, e.g., Ref. \cite{PS,SP}, the action is
\begin{equation}
S = \sum_y \chibar(y) [\gam\mu D_\mu + \iden m] \chi(y) \ ,
\end{equation}
and is invariant under
translations, rotations, spatial inversions and charge conjugation.
When $m\to0$, it is also invariant under the axial transformations
\[
\chi    \to         \exp[i\alpha\sfno55] \chi\ , \
\chibar \to \chibar \exp[i\alpha\sfno55] \ .
\]

In order to improve the action, one adds operators of $d=5$,
with coefficients adjusted order by order in perturbation theory so as to
cancel $O(a)$ terms in correlation functions.
(In fact, except in scalar theories, only on-shell quantities 
are improved.)
That this improves all on-shell quantities at once
is the non-trivial assumption, shown by Symanzik for scalar theories.
The $d=5$ operators must have the same
symmetries as those in the original action.
Ignoring axial symmetry, the allowed operators are \cite{sharpeprep}
\begin{equation}
\nonumber
m^2 \chibar \iden \chi \ ,\ \ \chibar \iden D_\mu^2 \chi \ ,\ \
m \chibar \gam\mu D_\mu \chi \ , \ \
\chibar \gam{\mu\nu} F_{\mu\nu} \chi \ .
\end{equation}
%For example, translation invariance allows only flavor singlets.
However, none of these operators is consistent with the axial
symmetry---treating $m$ as a spurion field one can show
that the bilinear must either contain an even number of links,
and be multiplied by an odd function of $m$, 
or contain an odd number of links and be multiplied by an even function of $m$.
Since there are no operators available to improve the action, 
it must already be good to $O(a^2)$.

Now I proceed to $B_K$. I will discuss the matrix
element of the four-fermion operator $\co_B$
in the numerator of Eq. \ref{bkeqn};
a similar argument works for the simpler matrix elements in the denominator.
To improve a matrix element one must not only improve the action, but also
improve the operator itself. Since the staggered action is already improved,
one needs only consider the operator.
I assume, following the second paper in Ref. \cite{symanzik}, that
improvement can be accomplished for all on-shell matrix elements by adding
$d=6$ and 7 operators to the original operator, 
with coefficients determined order by order in perturbation theory.
(Mixing with lower dimension operators is forbidden by the flavor structure.)
I assume further that these operators must have the same symmetries as
the original operators.

Let $\co^6_\CONT$ be a vector of continuum $d=6$ operators.
At tree-level these match onto a corresponding vector of
lattice operators $\co^6_\LATT$. 
These lattice operators will mix with
all others of $d=6$ having the same symmetries, 
so one must extend the vector $\co^6_\LATT$
to include all the possibilities.
I assume that the continuum vector is similarly extended.
Let $\co^7_\LATT$ be a vector containing all operators
with $d=7$ at tree-level having the same symmetries as the $\co^6_\LATT$.
The assumption of all-orders improvement is then
\begin{equation}
\co^6_\CONT (1 + O(a^2)) = c(g^2) \co^6_\LATT + d(g^2) a \co^7_\LATT \ ,
\end{equation}
in the sense that lattice matrix elements of the operator on the
r.h.s. equal those of the continuum operators up to corrections of $O(a^2)$.
The coefficients $c$ and $d$ are matrices.
$c$ is square, and is the identity at tree-level.
$d$ is rectangular, and begins at $O(1)$ in general.

We do not want to have to calculate $c$ and $d$ to all orders. Thus we
want an expression for $\co^6_\LATT$ in terms of continuum matrix elements
alone. This requires relating $\co^7_\LATT$ to the corresponding
continuum operators $\co^7_\CONT$. The assumed form is
\begin{equation}
a \co^7_\CONT (1 + O(a)) = \widetilde{c}(g^2) \co^6_\LATT 
+ \widetilde{d}(g^2) a \co^7_\LATT  \ ,
\end{equation}
where $\widetilde{d}$ is a square matrix which is the identity at tree-level.
$\widetilde{c}=O(g^2)$ represents the fact that
$d=7$ operators mix back into $d=6$ operators.

Reorganizing these equations we find 
\begin{equation}
\co^6_\LATT = (1 + O(a^2)) 
(c - d {\widetilde d}^{-1} {\widetilde c})^{-1} 
\left( \co^6_\CONT - d {\widetilde d}^{-1} a \co^7_\CONT \right) \ .
\end{equation}
This shows that the $O(a)$ terms in matrix elements of
$\co^6_\LATT$ can be obtained from the {\em continuum} matrix elements
of $\co^7_\CONT$.
The perturbative matrix on the r.h.s. is of the form $1+O(g^2)$---
multiplying by its inverse calculated to, say, 1-loop,
removes $O(g^2)$ corrections to the matching between continuum
and lattice matrix elements.

I now apply this equation to $\co_B$.
One must transcribe the continuum operator onto the lattice,
and then write down the (long) list of operators $\co^6_\LATT$
and $\co^7_\LATT$. Various choices of lattice operator,
all agreeing at tree-level, have been used.
We use Landau-gauge operators without gauge links on $2^4$ hypercubes, 
and smeared Landau-gauge operators on $4^4$ hypercubes \cite{PS}
(these have $d=O(g^2)$).
Gauge-invariant operators have been used in Ref. \cite{ishilat92prl}.
The argument is the same for all these choices, because they behave the
same way under the relevant symmetries: the hypercubic cube,
and the separate axial rotations of the four fermion fields.

The crucial point is this.
With staggered fermions the continuum theory has four degenerate
versions of each quark, and a corresponding flavor symmetry.
The continuum operators of interest have flavor $\xi_5\times\xi_5$.
It turns out, however, that none of the 
$d=7$ operators has this flavor \cite{sharpeprep}.
Thus, if we take the matrix elements between a $K$ and $\overline{K}$
both of flavor $\xi_5$, 
the contributions of $\co^7_\CONT$ {\em vanish identically}.
Thus there are no $O(a)$ corrections to these particular matrix elements:
they are automatically improved.

There are other operators in $\co^6_\CONT$ having flavor $\xi_5\times\xi_5$,
but these are multiplied by coefficients of $O(g^2)$ ($O(g^4)$ if one uses
one-loop matching).

A concern with this argument is that 
$c$ and $\widetilde{c}$ might contain non-perturbative 
parts of $O(a)$. This will not matter, however,
as long as the symmetry properties are retained.

\section{OTHER IMPROVEMENTS}

We use the same set of lattices as in Ref. \cite{sharpelat91}, 
but we now have results with two sets of operators:
Landau-gauge unsmeared and smeared.
Both have $O(a^2)$ scaling corrections, but they should agree
in the continuum limit. This provides a consistency check. 

We have now included one-loop perturbative corrections.
Patel and I have calculated these for both the original
and smeared operators\cite{PS,SP}, 
the former results being in agreement with those of Ref. \cite{ishishiz}.
The results are of the form
\begin{equation}
% for unsmeared (tadpole improved)
\co_B^\CONT(\NDR,\mu) =
\left[1 + {g^2\over 16\pi^2}4\ln({\pi\over a\mu e^{1/3}})\right]\co_B^\LATT 
+ {g^2\over 16\pi^2} \delta\co^\LATT \ .
\label{bkperteqn}
\end{equation}
To extrapolate to the continuum, we choose a fixed scale: $\mu=2$ GeV.

Lepage and Mackenzie have shown that perturbative corrections are reliable
if one uses the correct expansion parameter \cite{lepamack}.
For the coefficient of $\delta\co$, we use $g^2$ determined
from $\Tr(U)$ in Landau gauge, which yields
$g_U^2=1.82,1.66,1.54$ for $\beta=6,6.2,6.4$.
For the coefficient of logarithm, which represents the effect of loop
momenta between $\pi/a$ and $\mu$, we use either $g_U^2$,
or the value obtained by running from $g_U^2=1.82$ at $\beta=6$ to $\mu=2$GeV
using the 2-loop $N_f=0$ formula, assuming that the starting scale is
$\pi/a$. For $1/a=1.9$GeV at $\beta=6$, this gives 
$g_U^2(2{\rm GeV})=2.72$.
In the end we take the average of these two methods, and use half the
difference as an estimate of the systematic error.
We are in the process of reducing this uncertainty
using the automatic scale fixing procedure of Ref. \cite{lepamack}.
The perturbative corrections are small for unsmeared operators,
increasing the final result by $\sim 3\%$. 
The corrections are larger (up to 10\%) for smeared operators.

Although $B_K$ is dimensionless, it has a weak dependence on
the lattice spacing because of the
anomalous dimension factor in Eq. \ref{bkperteqn}, 
and through the value of the lattice kaon mass.
We use updated values of $a$ determined from $m_\rho$:
$1/a=1.9, 2.5, 3.55$ GeV for $\beta=6,6.2,6.4$.
Repeating the analysis using $a$ determined from $f_\pi$ gives an estimate
of the corresponding systematic error.

For $\beta=6$ our bare lattice numbers are unchanged from those
presented in Ref. \cite{sharpelat91}, 
and since confirmed by Ref. \cite{ishilat92prl}.
At $\beta=6.2$ and $6.4$ our calculation is hampered by the relative
shortness of our lattices in the time direction, which leads to
contamination from more massive states, particularly $\rho$ mesons.
We use two methods of calculation
each with different sources of contamination \cite{sharpelat90}.
We now understand how to subtract these contaminations using
the data itself \cite{usinprep}.
To be conservative, we use the size of the subtractions as
an estimate of the systematic error.

\begin{figure}[t]
%\vspace{-0.2truein}
%\centerline{\psfig{file=bkvsa.ps,width=2.8truein}}
\centerline{\psfig{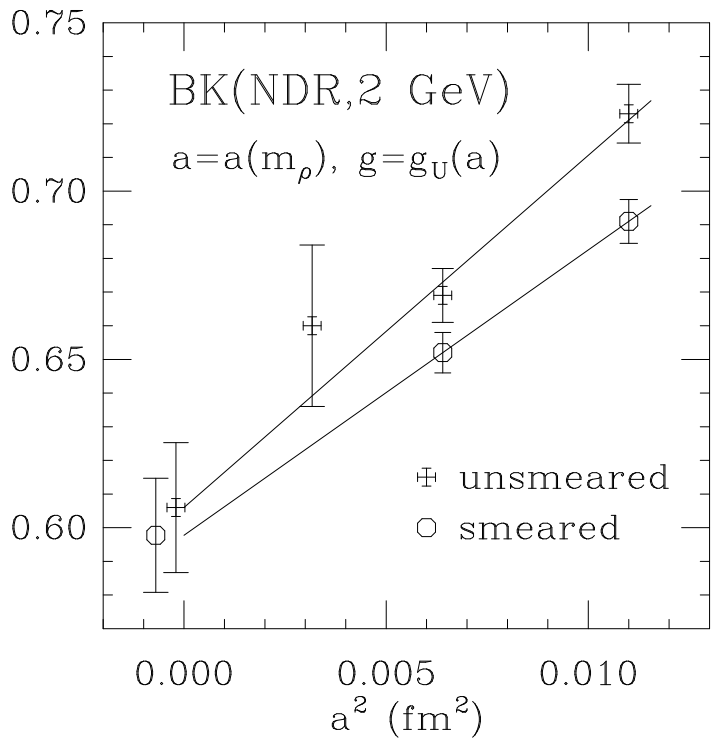}}
\vspace{-0.2truein}
\caption{Extrapolating $B_K$ to $a=0$.}
\vspace{-0.2truein}
\label{fig:bkvsa}
\end{figure}

An example of the extrapolation to $a=0$ is shown in Fig. \ref{fig:bkvsa}.
The data is not good enough to distinguish between linear and quadratic
dependence on $a$, although it favors the latter.
Thus we rely on the theoretical argument given above and assume a
quadratic dependence.
%The result at $\beta=6.4$ (only available for the unsmeared
%operator) has little effect on the extrapolation because of its large errors.
An important consistency check is that smeared and unsmeared operators
agree in the continuum limit.
It turns out that they also have similar dependence on $a$.
This is only true, however, after inclusion of perturbative corrections.
It has been found in Ref. \cite{ishilat92prl} that gauge invariant operators
also give consistent results, including the $a$ dependence,
once perturbative corrections are included.
Thus the relatively large scaling violations do {\em not} appear 
to be an artifact of using gauge non-invariant operators.
For our central value we use the average of
the extrapolated results from unsmeared and smeared operators,
and take half the difference as an estimate of a systematic error.
For the statistical error we take the larger of the two errors.

\section{NUMERICAL RESULTS}

The preliminary result from our analysis is,
in the quenched approximation and for degenerate quarks,
\begin{eqnarray*}
B_K(\NDR, 2{\rm GeV}) &=& 0.616 \pm 0.020 (stat) \\
&& \mbox{}\pm 0.014 (g^2) \\
&& \mbox{}\pm 0.009 (scale) \\
&& \mbox{}\pm 0.004 (operator) \\
&& \mbox{}\pm 0.002 (contamination) \\
&=&0.616 \pm 0.020 \pm 0.017 \ ,
\end{eqnarray*}
where, in the last line, we have combined all the systematic errors
in quadrature. 
%This result can be combined directly 
%with tabulated values of coefficient functions at $\mu=2$GeV.
It is more conventional to quote a result for the
scale independent B-parameter, $\bkhat$.
Using the continuum $\alpha_s$ evaluated at 2 GeV 
with $\Lambda^{(4)}_{\overline{\rm MS}}=300$MeV,
and the continuum anomalous dimension, we find
\begin{eqnarray}
\nonumber
\bkhat &\equiv& B_K(\NDR, 2{\rm GeV})\ \alpha_s(2 {\rm GeV})^{-6/25} \\
&=& 0.825 \pm 0.027 \pm 0.023 \ .
\end{eqnarray}
The major change from Ref. \cite{sharpelat91} is the
use of quadratic extrapolation.
Perturbative corrections also increase the result, by $\sim 3\%$,
and a similar increase results from 
the use of a different (and better) definition of $\bkhat$.

using a different (and better)
method of matching to the continuum $\bkhat$.

Errors due to quenching and to the use of degenerate quarks
are {\em not} included in these results.
There are reasons to think, however, that these errors are 
comparable to those quoted above.
Unquenched calculations are now possible 
for quark masses $m_q\sim m_s/2$,
on lattices with spacing $1/a\sim 2$GeV, and find
results for $B_K$ which agree within errors with quenched results
\cite{ishilat92prl,kilcupprl}.
This is surprising and encouraging.
It must be tested at smaller lattice spacings to determine
whether the full and quenched $a$ dependences are similar.

In most quantities a more important issue would be the dependence
on the light quark masses. For $B_K$, however, the dependence enters
at non-leading order. If one uses the chiral logarithms to estimate
the order of magnitude of the correction, 
one finds a 3\% increase for non-degenerate quarks 
\cite{sharpechlog,usinprep}. It is important to check this
with unquenched simulations for $m_s\ne m_d$.
Quenched data is a not good guide
because of contamination from $\eta'$ loops \cite{sharpechlog}. 

If the result for $\bkhat$ withstands further scrutiny,
it will have considerable phenomenological impact.

\section*{ACKNOWLEDGEMENTS}
I would like to thank Claude Bernard, Rajan Gupta,
Greg Kilcup, Paul Mackenzie, Apoorva Patel
and Akira Ukawa for useful conversations.
This work is supported in part by the DOE through
grant DE-FG06-91ER40614, and by an Alfred P. Sloan Fellowship.
%

%%%%%%%%%%%%%%%%%%%%%%%%
% journal and conference list
%

\def\PRL#1#2#3{{Phys. Rev. Lett.} {\bf #1}, #3 (#2) }
\def\PRD#1#2#3{{Phys. Rev.} {\bf D#1}, #3 (#2)}
\def\PLB#1#2#3{{Phys. Lett.} {\bf #1B} (#2) #3}
\def\NPB#1#2#3{{Nucl. Phys.} {\bf B#1} (#2) #3}
\def\NPBPS#1#2#3{{Nucl. Phys.} {\bf B ({Proc. Suppl.}) {#1}} (#2) #3}
\def\etal{{\em et al}}

\end{document}